\documentclass[letterpaper, 10 pt, conference]{ieeeconf}  % Comment this line out
\IEEEoverridecommandlockouts                              % This command is only
\usepackage{graphicx}
\usepackage{amsmath} % assumes amsmath package installed
\usepackage{rotating}
\usepackage{tabularray}
\usepackage{hyperref}
\usepackage{subcaption} % in your preamble

\title{\LARGE \bf
    C-EQ-ALINEA --
    Distributed, Coordinated, and Equitable Ramp Metering Strategy for Sustainable Freeway Operations
}

\author{Kevin Riehl, Omar Alami Badissi, Anastasios Kouvelas and Michail A. Makridis% <-this % stops a space
\thanks{This work was not supported by any organisation.}% <-this % stops a space
\thanks{K. Riehl, O.A. Badissi, A. Kouvelas, and M. Makridis are with Traffic Engineering Group, Institute for Transportation Planning and Systems,
        ETH Zürich, Stefano Franscini Platz 3, 8053 Zürich, Switzerland
        {\tt\small kriehl@ethz.ch, oalami@student.ethz.ch, kouvelas@ethz.ch, mmakridis@ethz.ch }.}%
 }

\begin{document}

\maketitle
\thispagestyle{empty}
\pagestyle{empty}

% %%%%%%%%%%%%%%%%%%%%%%%%%%%%%%%%%%%%%%%%%%%%%%%%%%%%%%%%%%%%%%%%%%%%%%%%%%%%%%%%
\begin{abstract}
Ramp metering is a widely deployed traffic management strategy for improving freeway efficiency, yet conventional approaches often lead to highly uneven delay distributions across on-ramps, undermining user acceptance and long-term sustainability. 
While existing fairness-aware ramp metering methods can mitigate such disparities, they typically rely on centralized optimization, detailed traffic models, or data-intensive learning frameworks, limiting their real-world applicability, particularly in networks operating legacy ALINEA-based systems. 
This paper proposes C-EQ-ALINEA, a decentralized, coordinated, and equity-aware extension of the classical ALINEA feedback controller. 
The approach introduces lightweight information exchange among neighbouring ramps, enabling local coordination that balances congestion impacts without centralized control, additional infrastructure, or complex optimization. 
C-EQ-ALINEA preserves the simplicity and robustness of ALINEA while explicitly addressing multiple notions of fairness, including Harsanyian, Egalitarian, Rawlsian, and Aristotelian perspectives. 
The method is evaluated in a calibrated 24-hour microsimulation of Amsterdam’s A10 ring road using SUMO. 
Results demonstrate that C-EQ-ALINEA substantially improves the equity of delay distributions across ramps and users, while maintaining (in several configurations surpassing) the efficiency of established coordinated strategies such as METALINE. These findings indicate that meaningful fairness gains can be achieved through minimal algorithmic extensions to widely deployed controllers, offering a practical and scalable pathway toward sustainable and socially acceptable freeway operations.
Open source implementation available on GitHub.
\end{abstract}

%%%%%%%%%%%%%%%%%%%%%%%%%%%%%%%%%%%%%%%%%%%%%%%%%%%%%%%%%%%%%%%%%%%%%%%%%%%%%%%%
\section{INTRODUCTION}

Sustainable and intelligent transportation systems increasingly rely on active traffic management to make more efficient use of existing infrastructure. Among these measures, ramp metering has proven to be one of the most effective tools for stabilizing freeway flow, improving safety, and reducing congestion without costly physical expansion. As a result, ramp metering is widely deployed in urban freeway networks and remains a cornerstone of network-level traffic optimisation~\cite{zhang_levinson_2006}.

%Sustainable and intelligent transportation systems increasingly rely on active traffic management to maximise the efficiency of existing infrastructure. Among these strategies, ramp metering remains one of the most effective tools for stabilizing freeway flow, improving safety, and reducing congestion without costly expansion, and is widely deployed in urban networks~\cite{zhang_levinson_2006}.

However, sustainability in traffic management extends beyond efficiency alone. Control strategies that are perceived as unfair or excessively burdensome to specific user groups cause low user compliance through signal violation, behavioural adaptation, and public opposition against implementation and initiatives for revoking of existing implementations, ultimately undermining their long-term effectiveness~\cite{riehl2024towards,zhang_levinson_2006}. 
From this perspective, fairness is not only a social objective but a prerequisite for sustainable deployment of intelligent transportation systems~\cite{riehl2024quantitative}.
This challenge is particularly pronounced in the context of ramp metering. 

%Yet sustainability extends beyond efficiency. Control strategies perceived as unfair can trigger low compliance, behavioural adaptation, and public opposition, ultimately undermining long‑term effectiveness~\cite{riehl2024towards}. Fairness is therefore not only a social goal but a prerequisite for sustainable deployment of intelligent transportation systems~\cite{riehl2024quantitative}.

ALINEA~\cite{Papageorgiou1991} -- a classical local feedback controller - regulates on-ramp inflow based on downstream traffic conditions and has demonstrated robust efficiency gains. Yet, by treating each ramp in isolation, such approaches often lead to highly uneven distributions of waiting times, where users at certain ramps experience persistent and excessive delays. 
Empirical evidence has shown that these disparities can trigger signal violations, queue bypassing, and strong public resistance, including organised initiatives to deactivate or revoke ramp metering systems altogether. In several documented cases, political intervention followed precisely because the system was perceived as unfair, despite its aggregate efficiency benefits.

%The challenge is particularly evident in ramp metering. While the classical ALINEA controller~\cite{Papageorgiou1991} achieves robust efficiency by regulating on‑ramp inflow based on downstream conditions, its isolated operation often produces unequal waiting times, with some ramps experiencing persistent excessive delays. Such disparities have led to signal violations, queue bypassing, and in several cases, political interventions to revoke metering systems perceived as unfair, despite their overall performance benefits.

To address related issues to isolated ramp metering, prior research has highlighted the value of coordinated and fairness-aware ramp metering strategies. Centralised and hierarchical approaches such as METALINE~\cite{Papageorgiou1990meta}, HERO~\cite{Papamichail2010} or SWARM~\cite{Kotsialos2004}, and model predictive control explicitly consider interactions between ramps and can improve both efficiency and equity~\cite{zhang_levinson_2005}. 
Furthermore, a growing branch of research is dedicated to incorporating fairness metrics directly into optimisation objectives or employ learning-based methods to balance efficiency and equity. 
While effective in theory and simulation, these methods typically rely on centralised computation, extensive calibration, detailed traffic models, or large data requirements, which limit their scalability and real-world applicability -- particularly for networks already operating legacy ALINEA-based systems.

%Research on coordinated and fairness‑aware metering aims to mitigate these issues. Centralised and hierarchical methods -- such as METALINE~\cite{Papageorgiou1990meta}, HERO~\cite{Papamichail2010}, and SWARM~\cite{Kotsialos2004} -- explicitly consider interactions between ramps to improve efficiency and equity. 
%More recent learning‑based approaches embed fairness metrics in their objectives~\cite{zhang_levinson_2005}. 
%However, such methods typically depend on centralised computation, detailed models, and extensive calibration or data requirements, limiting their scalability and adoption in networks already operating legacy ALINEA‑based systems.

This study proposes C‑EQ‑ALINEA -- an equity‑aware, decentralised, and coordinated extension of the ALINEA ramp‑metering algorithm. 
It introduces limited information exchange among neighbouring ramps, enabling local coordination that balances congestion impacts without centralised control or additional infrastructure. Evaluated in SUMO microsimulation on Amsterdam’s A10 ring road, C‑EQ‑ALINEA achieves a more equitable distribution of delays across ramps while maintaining efficiency comparable to established coordinated methods. 
By incorporating simple modifications to one of the most widely used ramp‑metering controllers, the approach promotes behavioural, institutional, and operational fairness, offering a practical pathway for real‑world deployment.

A reproducible open‑source implementation is available at \url{https://github.com/DerKevinRiehl/c_eq_alinea}.

%%%%%%%%%%%%%%%%%%%%%%%%%%%%%%%%%%%%%%%%%%%%%%%%%%%%%%%%%%%%%%%%%%%%%%%%%%%%%%%%
\section{RELATED WORKS}

The tension between efficiency and fairness in ramp metering has motivated a range of control strategies that explicitly incorporate equity considerations into their decision-making logic~\cite{riehl2024towards}.

\begin{table*}[!ht]
    \caption{\textbf{Fairness-Oriented Ramp Metering Algorithms.}}
    \label{tab:rm_fairness}
    \centering
    
    \begin{tabular}{p{2.5cm} p{3.2cm} p{3.2cm} p{3.2cm} p{3.2cm}}
        \hline
        \textbf{Algorithm} & \textbf{Objective} & \textbf{Method} & \textbf{Strengths} & \textbf{Limitations} \\
        \hline
        AMOC (2001)~\cite{kotsialos2001efficiency}  & Delay vs. ramp storage trade-off & Storage-aware optimal control & Corridor-wide fairness & Sensitive to config \\
        BEEN (2005)~\cite{zhang_levinson_2005} & Efficiency-fairness balance & Weighted travel time minimization & Dynamic fairness prioritization & Sensitive to weights \\
        FTRM (2013)~\cite{kesten_2013} & Static equity in fixed-time & Multi-index cycle time optimization & Equity metrics: Gini, Theil, etc. & No reactivity \\
        Pareto Optimization (2016)~\cite{ling2016pareto} & Equity vs. efficiency trade-off & Multi-objective optimization (MCTM) & Captures trade-off frontier & High computation cost \\
        Multi-Class Sustainable (2017)~\cite{pasquale2017sustainable} & Inter-vehicle fairness + environment & Dual-class feedback control & Handles emissions + truck fairness & Early stage, limited testing \\
        Bargaining Game (2021)~\cite{bargaining_game_2021} & Distributed fairness & Game-theoretic MPC & No ramp dominates delay & Requires coordination infra \\
        Fair RL (2021–2022)~\cite{han2021ramp,Han2022} & Equity-aware learning & RL with fairness constraints & Learns balance between users & Data intensive \\
        \hline
    \end{tabular}
\end{table*}

\textit{AMOC}~\cite{kotsialos2001efficiency} formulates ramp metering as a corridor-wide optimal control problem in which ramp storage constraints are used to implicitly balance efficiency and equity. While effective in coordinating multiple ramps, AMOC relies on centralised optimisation and detailed traffic models, limiting its scalability and real-time applicability.

\textit{BEEN}~\cite{zhang_levinson_2005} introduces fairness by minimising weighted travel time, where ramp delays are dynamically penalised as waiting times increase. Although this approach improves equity in delay distribution, it requires careful weight calibration and operates within an optimisation framework that is not easily transferable to existing local controllers.

\textit{Fixed-Time Ramp Metering (FRTM)}~\cite{zhang_levinson_2005} incorporates multiple equity indices, such as Gini and Theil measures, to optimise cycle times under static demand assumptions. Despite offering explicit fairness guarantees, the lack of real-time responsiveness constrains its effectiveness under fluctuating traffic conditions.

\textit{Pareto-based optimisation}~\cite{Qiong_2011} frames ramp metering as a multi-objective problem, explicitly exploring the trade-off frontier between efficiency and equity. While conceptually rigorous, such approaches require global network knowledge and incur significant computational overhead, rendering them impractical for operational deployment.

\textit{Multi-class sustainable ramp metering}~\cite{pasquale2017sustainable} extends feedback control to account for heterogeneous vehicle classes, addressing both environmental and inter-class fairness objectives. However, the approach increases model complexity and depends on detailed vehicle classification and emission modelling.

\textit{Bargaining game-based methods}~\cite{bargaining_game_2021} model ramp controllers as strategic agents that negotiate metering rates to prevent disproportionate delays. Although distributed in nature, these methods require iterative coordination and reliable communication, raising concerns regarding convergence, robustness, and real-time feasibility.

\textit{Fair reinforcement learning (RL)} approaches~\cite{han2021ramp,Han2022} embed equity metrics into learning-based control policies, enabling adaptive balancing of efficiency and fairness. Despite their flexibility, RL-based solutions are data-intensive, difficult to interpret, and currently lack the transparency and reliability required for safety-critical traffic control.

The comparative review of prior works on fairness-oriented ramp metering~\cite{kotsialos2001efficiency,zhang_levinson_2005,Qiong_2011,pasquale2017sustainable,bargaining_game_2021,han2021ramp,Han2022} in Table~\ref{tab:rm_fairness} reveals three main limitations: 
(i) reliance on centralised optimisation, computationally-intensive, or data-hungry learning frameworks, which restrict scalability, real-world and real-time applicability;  
(ii) dependence on global models or calibrated weights rather than embedded feedback for fairness enforcement; and  
(iii) limited exploration of fairness notions, as most studies adopt a primarily Egalitarian notion while overlooking the multidimensional character of equity.

% Across these approaches, fairness is predominantly achieved through centralised optimisation, explicit multi-objective formulations, or computationally intensive learning frameworks. 
% While effective in controlled settings, such solutions often require extensive calibration, detailed models, or high-quality data, which limits their applicability to real-world freeway networks already operating local feedback controllers.

C-EQ-ALINEA addresses fairness through lightweight distributed coordination, extending the well-established ALINEA structure via minimal information exchange among neighbouring ramps. 
By avoiding centralised optimisation and preserving local feedback control, the proposed approach offers a practical and scalable pathway to equitable and therefore sustainable ramp metering.

%%%%%%%%%%%%%%%%%%%%%%%%%%%%%%%%%%%%%%%%%%%%%%%%%%%%%%%%%%%%%%%%%%%%%%%%%%%%%%%%
\section{METHODS}

\subsection{Notation \& Problem Statement}

The highway is modelled as a directed graph $\mathcal{G}=\{N,Z\}$, where nodes $N$ represent points on the highway where on-ramps (and off-ramps) enable vehicles to enter (and leave) the highway links $Z$.
At each on-ramp $n \in N$, a traffic signal controls the vehicle inflow rate $q_n$. 
The ramp-metering controller regulates $q_n[k]$ at control cycle $k$ based on real-time measurements obtained from induction loop detectors installed along the mainline, typically spaced 100 - 200m apart downstream. 
These sensors provide occupancy data $o_n{k}$ that serve as a proxy for traffic density, enabling responsive and adaptive ramp metering.
The symmetric on-ramp neighbourhood of nodes of $n$ is denoted with $\mathcal{N}[n]$, including $m$ on-ramp nodes before (upstream) and after (downstream) node $n$.

The controller applies the bounded metering rate $r_n[k]$ (share of time in cycle $k$ where vehicles can pass, between 0 and 100\%), that can be determined assuming an average discharge rate $\gamma$ (usually 0.5 s/veh) and a cycle duration $D$ as follows:
\begin{equation}
    r_n[k] = (q_n[k] \times \gamma) / D
\end{equation}

\subsection{ALINEA}

The feedback controller ALINEA~\cite{HadjSalem1990} updates the metering rate according to a simple yet powerful P-control law given a critical, desired occupancy $\hat{o}$, and an control gain factor $K$ as follows:
\begin{equation}
    q_n[k] = q_n[k-1] + K \times (\hat{o} - o_n[k])
\end{equation}

\subsection{C-EQ-ALINEA}

C-EQ-ALINEA extends this local feedback control by incorporating flow information from neighboring ramps to achieve better network-wide coordination. 
The algorithm maintains the simplicity of ALINEA, while adding a coordination term $q_n^{coord}[k]$ enabling decentralised yet coordinated ramp metering:

\begin{equation}
    q_n[k] = q_n[k-1] + K \times (\hat{o} - o_n[k]) + q_n^{coord}[k]
\end{equation}

The coordination term incorporates a coordination gain factor $K_c$, and the difference between local $q_n[k]$ and weighted neighbouring flow rates $w_j \times q_j[k] \; \forall \; j \in \mathcal{N}[n]$ with $\sum_j w_j = 1$ as follows:

\begin{equation}
    q_n^{coord}[k] = K_c \times (\sum_{j\in\mathcal{N}[n]} w_j q_j[k] - q_n[k])
\end{equation}

The un-normed weights $u_j$ are calculated using the distance $d_{nj}$ of $n$ to its neighbour $j$ as follows:
\begin{equation}
    u_j = max(0, 1 - \frac{d_{nj}}{L_{max}})
\end{equation}

The weights $w_j$ are calculated using $u_j$ as follows:
\begin{equation}
    w_j = u_j / \sum_j(u_j)
\end{equation}

In this work we explore two approaches to weight calculation, using global distance normalisation ($L_{max}$ represents maximum distance across all consecutive ramp pairs in the network) and local distance normalisation ($L_{max}$ represents maximum distance among all consecutive ramp pairs in the neighbourhood).

%%%%%%%%%%%%%%%%%%%%%%%%%%%%%%%%%%%%%%%%%%%%%%%%%%%%%%%%%%%%%%%%%%%%%%%%%%%%%%%%

\begin{figure*}[!ht]
    \centering
    \includegraphics[width=\linewidth]{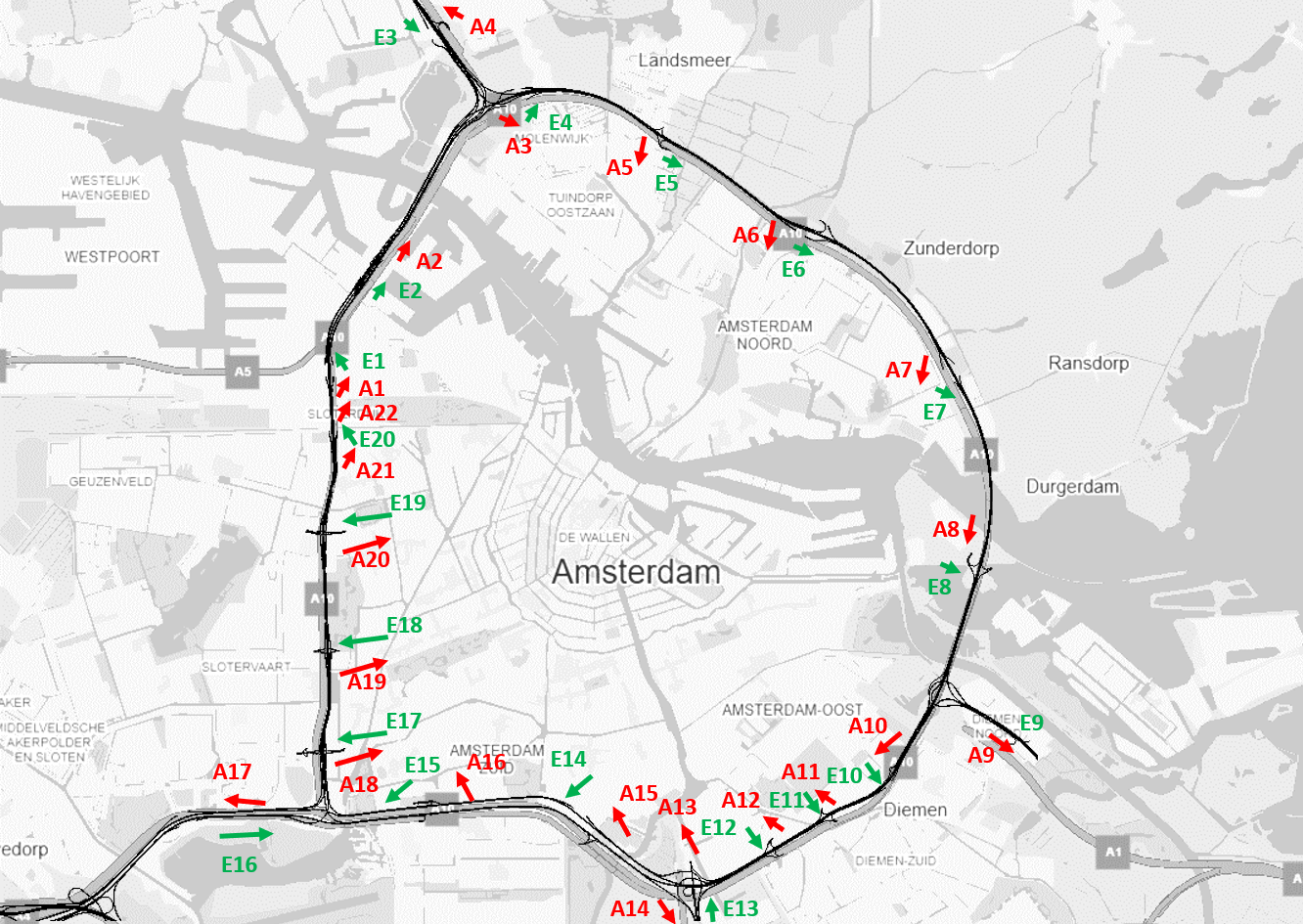}
    \caption{\textbf{Case Study Network: Amsterdam A10 Ring-Road (Netherlands).}}
    \label{fig:placeholder}
\end{figure*}

\section{RESULTS}
The proposed controllers were benchmarked using a calibrated, real‑world traffic microsimulation of the A10 ring-road highway network in Netherlands (see Fig.~\ref{fig:placeholder}). 
The 32km long beltway around Amsterdam comprises 20 on-ramps and 22 off-ramps. 
The demand model was calibrated with data provided by national administrations (\textit{Nationaal Dataportaal Wegverkeer}).
Simulations were running for 24h (including an additional 3000s warmup period) and were conducted in the open‑source environment SUMO with a mixed, multimodal vehicle fleet composition derived from national statistics (\textit{Centraal Bureau voor de Statistiek}). 
All reported results represent the mean values across 10 simulations with different random seeds, with standard deviations in brackets, where the highly congested period from 10:00 AM to 06:00 PM was used for evaluation. 
For more details, please find the GitHub repository.

\subsection{Efficiency Analysis}

\begin{sidewaysfigure}
    \centering
    \begin{minipage}{0.32\linewidth}
        \centering
        \subcaptionbox{No control\label{fig:no_control}}{%
            \includegraphics[width=\linewidth]{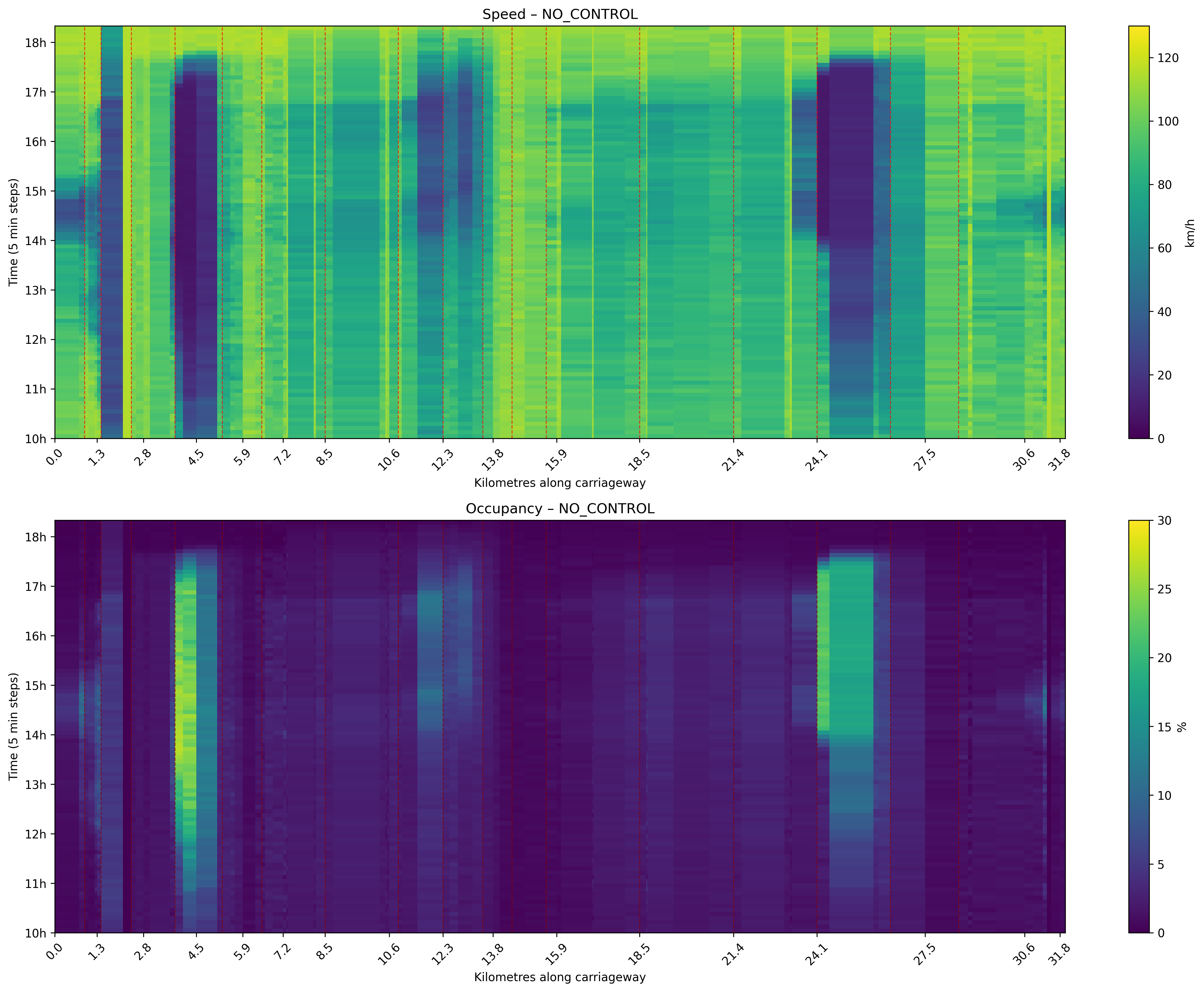}}
    \end{minipage}%
    \begin{minipage}{0.32\linewidth}
        \centering
        \subcaptionbox{ALINEA\label{fig:alinea}}{%
            \includegraphics[width=\linewidth]{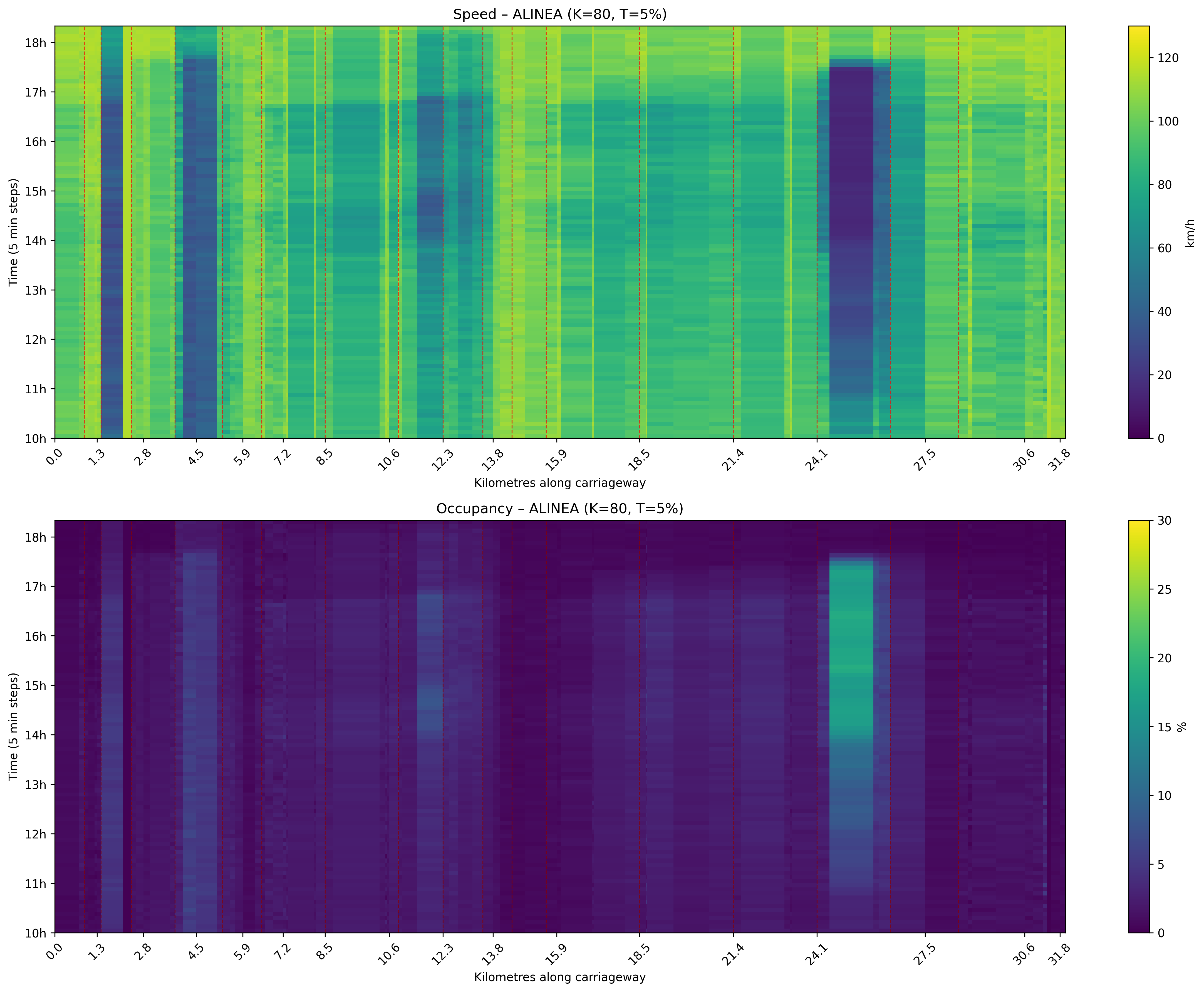}}
    \end{minipage}%
    \begin{minipage}{0.32\linewidth}
        \centering
        \subcaptionbox{METALINE\label{fig:metaline}}{%
            \includegraphics[width=\linewidth]{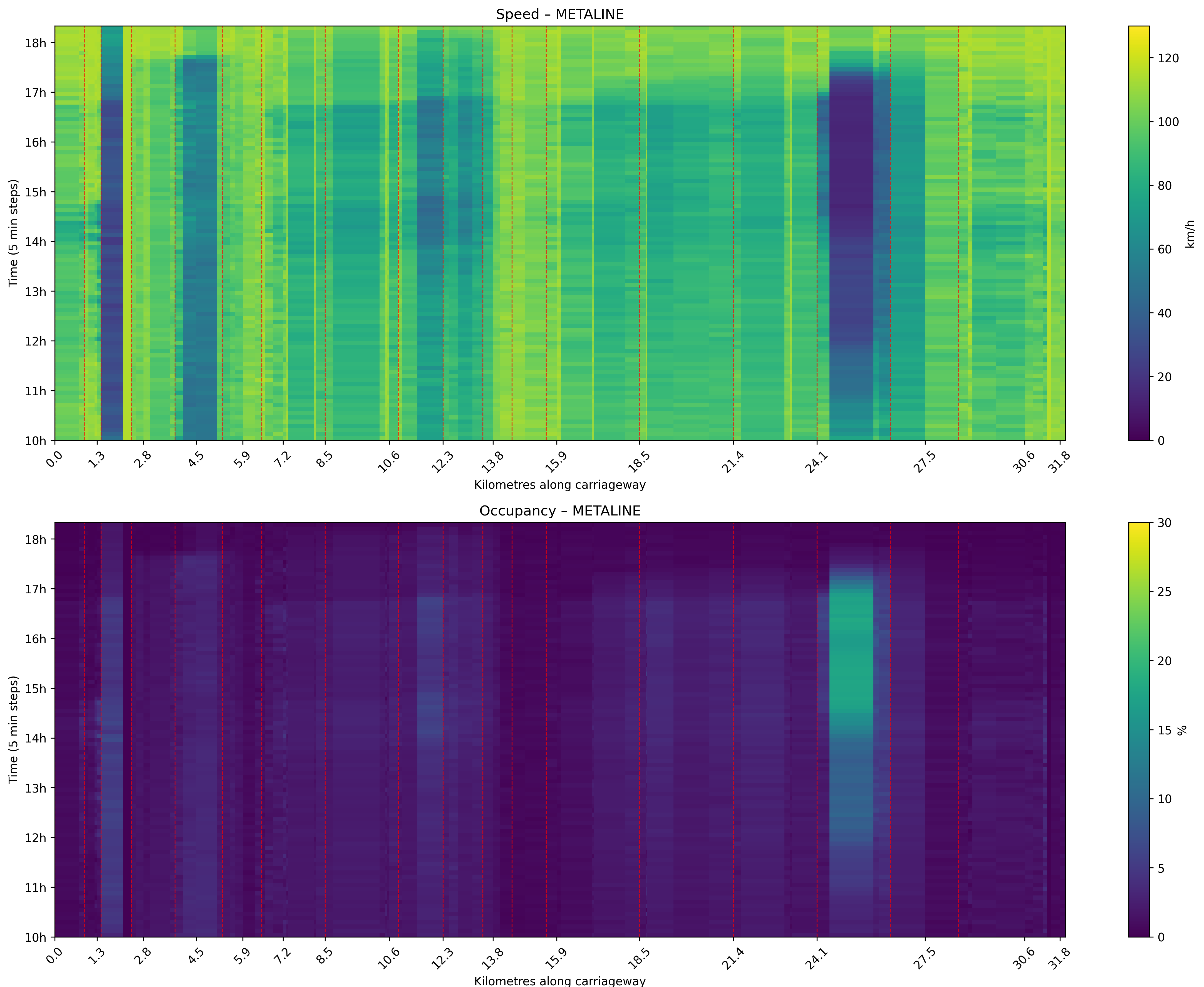}}
    \end{minipage}

    \caption{\textbf{Efficiency Analysis: Ramp Metering.}}
    \label{fig:placeholder2}
\end{sidewaysfigure}

To assess the efficiency of C-EQ-ALINEA, it is compared with two baseline controllers -- ALINEA and the coordinated METALINE -- to establish reference performance.
All controllers and their parameters were optimised based on an extensive grid-search with multiple random seeds per parameter combination to maximise network throughput (for exact parameters and implementation please refer to the GitHub repository). 
The results are shown in Table~\ref{tab:efficiency_analysis} and Fig.~\ref{fig:placeholder2}, where the baseline is considered to be the uncontrolled scenario.

\begin{table*}[!ht]
    \caption{\textbf{Efficiency Benchmark of Traffic Controllers.}}
    \label{tab:efficiency_analysis}
    \centering
    \begin{tabular}{lrrrrrrrrr}
        \hline
        \textbf{Metric} 
        & \textbf{No Control} 
        & \textbf{ALINEA} 
        & \textbf{METALINE}
        & \multicolumn{6}{c}{\textbf{C-EQ-ALINEA}} \\
         & & & &
        \multicolumn{3}{|c}{{Global Distance Norm.}} &
        \multicolumn{3}{c}{{Local Distance Norm.}} \\
        & & & & \multicolumn{1}{|c}{\textbf{$m=1$}} & \textbf{$m=2$} & \textbf{$m=3$} & \textbf{$m=1$} & \textbf{$m=2$} & \textbf{$m=3$} \\
        \hline
        \\
        {Total Departed Vehicles}& \textbf{44993.6} & 44811.5 & 43554.9 & 43749.1 & 42004.3 & 40797.9 & 43716.7 & 43795.4 & 42108.1 \\
        {Total Arrived Vehicles}& \textbf{44928.8} & 44657.6 & 43446.8 & 43612.2 & 41878.4 & 40677.8 & 43631.3 & 43659.3 & 41980.7 \\
        {Arrival Rate (\%)}       & \textbf{99.9}    & 99.7    & 99.8    & 99.7    & 99.7    & 99.7    & 99.8    & 99.7    & 99.7    \\
        {Total Travel Time (h)}   & 6717.1  & 5328.4  & 5077.0  & 4988.0  & 4527.1  &\textbf{ 4357.5}  & 5677.5  & 4982.8  & 4563.2  \\
        {Total Travel Distance (km)} & 315445.9 & 312641.4 & 302581.5 & 307921.7 & 297027.5 & \textbf{290486.6} & 310423.7 & 307919.9 & 297954.8 \\
        {Total Delay (h)}         & 3591.2  & 2230.1  & 2078.7  & 1936.3  & 1583.5  & \textbf{1478.8}  & 2601.3  & 1931.2  & 1610.4  \\
        {Average Speed (km/h)}    & 47.0  & 58.7    & 59.6    & 61.8    & 65.6    & \textbf{66.7}    & 54.7    & 61.9    & 65.3    \\
        {Average Delay (s/veh)} & 282 & 180 & 168 & 156 & 132 & \textbf{126} & 210 & 156 & 138 \\
        \\
        \hline
    \end{tabular}
\end{table*}

All control strategies deliver substantial efficiency gains compared to the uncontrolled scenario, while keeping the total number of departed and arrived vehicles close to baseline levels. This indicates that improvements are not achieved at the expense of throughput, but rather through more effective utilisation of available freeway capacity.

The classical ALINEA controller reduces total delay (TD) by approximately 37.9\%, from 3591.2~h to 2230.1~h, while increasing the average speed (AS) from 47.0~km/h to 58.7~km/h and lowering the average delay per vehicle (AD) from 4.7 to 3.0~min/veh. The total number of departed and arrived vehicles decreases only marginally, confirming that the observed efficiency gains are not driven by demand suppression.

METALINE achieves a further improvement in overall efficiency, reducing total delay to 2078.7~h (--42.1\%) and increasing average speed to 59.6~km/h, while maintaining vehicle counts close to the baseline. However, METALINE does not consistently outperform the best configurations of the proposed controller across all efficiency indicators.

C-EQ-ALINEA consistently outperforms standard ALINEA and, for several configurations, also exceeds METALINE in terms of efficiency. Using global distance normalisation, the strongest performance is obtained for a neighbourhood size of $m=3$, where total delay drops to 1478.8~h (approximately 58.8\% below the uncontrolled case), average speed increases to 66.7~km/h, and average delay is reduced to 2.1~min/veh. In parallel, total travel time decreases substantially from 6717.1~h to 4357.5~h, while both departed and arrived vehicle counts remain above 40\,000, indicating robust throughput preservation. Similar, though slightly less pronounced, improvements are observed for $m=2$, suggesting diminishing marginal gains beyond moderate neighbourhood sizes.

For local distance normalisation, performance becomes more sensitive to the choice of neighbourhood size $m$. While the configuration with $m=3$ still yields strong efficiency improvements -- reducing total delay to 1610.4~h (--55.2\%) and maintaining a high average speed of 65.3~km/h -- the $m=1$ case exhibits notably weaker performance, with a higher total travel time of 5677.5~h and smaller delay reductions. This sensitivity indicates that limited spatial awareness can restrict the effectiveness of purely local normalisation schemes under congested conditions.

Overall, C-EQ-ALINEA with global distance normalisation and $m=3$ emerges as the most effective configuration for the A10 network, offering the best combination of delay reduction, speed improvement, and throughput preservation. Importantly, these efficiency gains are achieved through lightweight distributed coordination rather than centralised optimisation, highlighting the practical potential of the proposed approach for real-world freeway operations.

Fig.~\ref{fig:placeholder2}(a) visualizes traffic congestion patterns in the form of speed and occupancy heatmaps over space and time. The red lines highlight the positions of controlled on-ramps. Fig.~\ref{fig:placeholder2}(b) and (c) demonstrate how ramp metering can effectively mitigate congestion. A visualization of C-EQ-ALINEA is not reported here for space reasons, as it resembles subfigures (b) and (c).

\subsection{Fairness Analysis}

To assess fairness, this section compares the average vehicle delay across different ramps to identify possible disparities in their treatment, as shown in Table~\ref{tab:fairness_avdelay}. 
Ramps with negligible demand were excluded from the analysis to avoid biasing the results. 
Four fairness notions are considered to capture the distributional effects of alternative control strategies: Harsanyian, Egalitarian, Rawlsian, and Aristotelian fairness. 
Harsanyian fairness~\cite{harsanyi1975can} evaluates the average outcome and is measured using the average delay. 
Egalitarian fairness~\cite{goppel2016handbuch} focuses on the dispersion of outcomes, quantified via the Gini coefficient of delays. 
Rawlsian fairness~\cite{rawls1971atheory} examines the situation of the worst-off and is measured by the maximum delay. 
Aristotelian fairness~\cite{riehl2024quantitative} considers proportionality between outcomes and contributions and is measured through the demand-weighted average delay per on-ramp.

\begin{table*}[!ht]
    \caption{Fairness Analysis: Situation Across Ramps (Average Delay in s)}
    \label{tab:fairness_avdelay}
    \centering
    \begin{tabular}{lrrrrrrrrr}
        \hline
        \textbf{Ramp} 
        & \textbf{No Control} 
        & \textbf{ALINEA}
        & \textbf{METALINE}
        & \multicolumn{6}{c}{\textbf{C-EQ-ALINEA}} \\
        & & & & \multicolumn{3}{|c}{Global Distance Norm.} & \multicolumn{3}{c}{Local Distance Norm.} \\
        & & & & \textbf{$m=1$} & \textbf{$m=2$} & \textbf{$m=3$} & \textbf{$m=1$} & \textbf{$m=2$} & \textbf{$m=3$} \\
        \hline
        \\
        \;\;\; A & 151.7 & 134.8 & 148.6 & 121.7 & 109.2 & 105.7 & 128.0 & 121.2 & 104.5 \\
        \;\;\; B & 402.0 & 136.9 & 101.6 & 152.4 & 101.0 & 88.7  & 318.2 & 157.3 & 103.7 \\
        \;\;\; C & 550.4 & 259.8 & 304.7 & 249.9 & 164.1 & 137.2 & 454.0 & 252.5 & 173.1 \\
        \;\;\; D & 188.8 & 145.4 & 150.7 & 137.3 & 115.5 & 111.7 & 240.2 & 135.2 & 117.8 \\
        \;\;\; E & 273.5 & 197.4 & 185.7 & 193.1 & 159.0 & 154.0 & 208.1 & 180.2 & 161.4 \\
        \;\;\; F & 235.3 & 179.5 & 176.9 & 143.2 & 133.4 & 121.5 & 256.2 & 146.5 & 130.1 \\
        \;\;\; G & 316.7 & 233.0 & 237.0 & 141.6 & 138.7 & 134.9 & 123.8 & 148.0 & 138.6 \\
        \;\;\; H & 282.7 & 208.2 & 204.0 & 113.1 & 115.1 & 115.7 & 102.1 & 120.1 & 119.6 \\
        \;\;\; I & 443.5 & 330.0 & 325.0 & 271.7 & 259.9 & 258.7 & 283.1 & 265.6 & 262.8 \\
        \;\;\; J & 66.7  & 47.7  & 90.2  & 53.1  & 48.9  & 67.1  & 54.9  & 49.0  & 70.4  \\
        \;\;\; K & 199.0 & 97.3  & 117.1 & 149.8 & 119.4 & 83.7  & 150.4 & 151.9 & 90.0  \\
        \\
        \hline
        \textbf{Notion of Fairness}\\
        \;\;\; Harsanyian (Avg.) & 282.8 & 179.1 & 185.6 & 157.0 & 133.1 & \textbf{125.4} & 210.8 & 157.0 & 133.8 \\
        %\;\;\; Egalitarian (Std.) & 133.1 & 75.2  & 73.9  & 58.7  & 49.8  & \textbf{48.6}  & 110.0 & 57.6  & 49.7  \\
        \;\;\; Egalitarian (Gini) & 0.2635 & 0.2354 & 0.2219 & 0.1993 & 0.1932 & \textbf{0.1892} & 0.2888 & 0.1936 & 0.1907 \\ 
        \;\;\; Rawlsian (Max.) & 550.4 & 330.0 & 325.0 & 271.7 & 259.9 & \textbf{258.7} & 283.1 & 265.6 & 262.8 \\
        \;\;\; Aristotelian (Weight.Avg.) & 289.4 & 174.3 & 169.6 & 161.9 & 129.4 & \textbf{118.2} & 242.4 & 165.7 & 131.0 \\
        \\
        \hline
    \end{tabular}
\end{table*}

% From an Harsanyian perspective, the average delay across all ramps is 282.8 seconds per vehicle in the uncontrolled scenario. 
% ALINEA and METALINE substantially improve this delay by 103.7 s (36.67\%) and 97.2 s (34.37\%).
% C-EQ-ALINEA is able to achieve even further decreases, and for global distance normalisation with $m=3$ neighbours it achieves a reduction of the delay by 157.4 s (55.66\%).
From a Harsanyian perspective, the average delay across all ramps is 282.8~s per vehicle in the uncontrolled scenario. 
ALINEA and METALINE substantially reduce this delay by 103.7~s (36.7\%) and 97.2~s (34.4\%), respectively. 
C-EQ-ALINEA achieves a further reduction, and under global distance normalisation with $m=3$ neighbours, the delay decreases by 157.4~s (55.7\%). 
% From an Egalitarian perspective, all control strategies remove the Gini coefficient towards more homogeneously distributed delays, were C-EQ-ALINEA achieves the most equal distribution of average delays across ramps (28.20\% reduction when compared with uncontrolled scenario).
From an Egalitarian perspective, all control strategies reduce the Gini coefficient, leading to more uniformly distributed delays. 
C-EQ-ALINEA attains the most equitable distribution, yielding a 28.2\% improvement compared with the uncontrolled baseline. 
% From an Rawlsian perspective, the worst-case situation (maximum delay) is consistently reduced through ramp metering, where C-EQ-ALINEA achieves the highest reduction (52.99\%). 
% From an Aristotelian perspective, the delays are distributed in a way that they reflect the contribution (demand) of each on-ramp to highway delay (are more proportional to those). Here, C-EQ-ALINEA achieves the highest alignment.
From a Rawlsian perspective, the worst-case (maximum) delay is consistently reduced by ramp metering, with C-EQ-ALINEA again providing the largest improvement (52.99\%). 
Finally, under the Aristotelian perspective, delay patterns align more closely with the respective demands of each on-ramp, demonstrating a fairer proportionality between contribution and experienced delay; C-EQ-ALINEA achieves the strongest alignment among all methods.

% In addition to this analysis, further metrics including queuing metrics (maximum, average, cumulative queue length, merging rates), metering statistics (number of phase changes, average metering rate), and user experience (maximum delays, waiting times) have been studied for an across-ramps fairness assessment. 
% These analyses yield insights similar to those reported in Table~\ref{tab:fairness_avdelay}.
Beyond these fairness perspectives, additional indicators -- including queuing metrics (maximum, average, and cumulative queue length, merging rates), metering statistics (phase change frequency and average metering rate), and user metrics (maximum delay, waiting time) -- were also compared across ramps. 
These results exhibit similar fairness trends to those summarised in Table~\ref{tab:fairness_avdelay}.

% Furthermore did we study the delay distribution across users based on their trip distance.
% This analysis revealed that in an uncontrolled scenarios the relative delays (delay per km travelled) are more un-equally distributed across travelled distances where longer distances had shorter relative delays, when compared with controlled scenarios. This was observed both for average and maximum relative delays.
% METALINE achieved the most equal distribution of these relative delays over travelled distances, while C-EQ-ALINEA (global distance normalisation and $m=3$) achieved the second most equal distribution.
Finally, an analysis of delay distributions across users stratified by trip distance reveals that in the uncontrolled scenario, relative delays (delay per kilometre travelled) are less evenly distributed, with longer trips typically facing lower relative delays. 
Controlled scenarios, by contrast, exhibit more equitable relative delays across trip lengths. 
Among all strategies, METALINE provides the most uniform distribution of relative delays across distances, with C-EQ-ALINEA (global normalisation, $m=3$) ranking second.

%%%%%%%%%%%%%%%%%%%%%%%%%%%%%%%%%%%%%%%%%%%%%%%%%%%%%%%%%%%%%%%%%%%%%%%%%%%%%%%%
\section{Discussion}

The results demonstrate that meaningful improvements in fairness can be achieved without sacrificing efficiency or relying on centralised optimisation. By introducing lightweight information exchange among neighbouring ramps, C-EQ-ALINEA consistently reduces extreme delays and improves the distribution of waiting times, while preserving throughput levels comparable to established coordinated strategies. This finding is particularly relevant in the context of sustainable traffic management, where long-term effectiveness depends not only on system efficiency but also on user acceptance and behavioural compliance.

From a sustainability perspective, the proposed approach contributes along three complementary dimensions. Behaviourally, reducing excessive and highly uneven ramp delays mitigates incentives for signal violations and queue bypassing, which are known to erode the effectiveness of ramp metering over time. Institutionally, a more equitable distribution of delays addresses one of the primary sources of public resistance against ramp metering systems, thereby increasing their political and societal viability. Operationally, the distributed coordination mechanism enables smoother network-wide responses to congestion without introducing additional infrastructure or computational overhead.

An important insight from the results is the sensitivity of fairness and efficiency outcomes to the spatial extent of coordination. While small neighbourhoods can already deliver improvements, larger neighbourhood sizes consistently yield more robust performance, particularly under global distance normalisation. This highlights a practical design trade-off between local responsiveness and corridor-level awareness, which can be tuned based on network topology and communication constraints.

Overall, C-EQ-ALINEA illustrates how simple algorithmic extensions to widely deployed controllers can yield substantial gains in fairness and sustainability, offering a pragmatic, real-world applicable pathway towards equitable and intelligent freeway operations.

%%%%%%%%%%%%%%%%%%%%%%%%%%%%%%%%%%%%%%%%%%%%%%%%%%%%%%%%%%%%%%%%%%%%%%%%%%%%%%%%
\section{Conclusions}

This paper set out to propose simple extension to enable decentralization, coordination, and fairness constraints of the most widely adopted ramp metering control system \textit{ALINEA} in the context of freeway networks.
The proposed method \textit{C-EQ-ALINEA} was evaluated based on a realistic, demand-calibrated microsimulation on Amsterdam's A10 ring road (Netherlands).
The benchmark results highlight significant fairness gains that could be achieved without severe sacrifices in efficiency, from a multitude of fairness definitions.

These findings have significant implications for practical freeway management.
C-EQ-ALINEA offers transportation agencies a deployable solution that can deliver
substantial congestion relief and improved network equity without requiring extensive computational infrastructure or complex optimization solvers. The algorithm's feedback structure ensures robustness to measurement noise and model uncertainties, critical factors for real-world implementation. 
If ALINEA-style loops and ramp meters already exist, C-EQ-ALINEA only needs modest add-ons: 
\begin{enumerate}
    \item communication infrastructure between ramps,
    \item simple firmware/software update so that each ramp controller can run the feedback logic
\end{enumerate}
No heavy new hardware or centralized optimizers are required, the local feedback design is already simple and robust.

\textbf{Limitations of this Work:}
\begin{itemize}
    \item Although multiple fairness notions were considered, perceived fairness and user acceptance were not directly measured and remain an important direction for future work.
    \item The evaluation is based on microsimulation of a single, albeit complex, urban freeway network. While the A10 ring road represents a demanding test case, further studies are required to assess transferability to different network structures, demand patterns, and incident scenarios. 
    \item C-EQ-ALINEA might suffer from incomplete detector coverage and aging infrastructure, which can reduce data quality.
    \item More over, it assumes sufficient on-ramp storage, but many ramps lack space, risking queue spillback onto surface streets.
    \item Currently, C-EQ-ALINEA does not handle physical constraints like limited ramp storage, which may reduce real-world effectiveness.
    \item Furthermore, performance may vary across different network layouts due to factors like ramp spacing, interchanges, managed lanes, and weaving sections.
\end{itemize}

\textbf{Ideas For Future Works:}
Future work should not only consider delays, but also address perceived fairness and user acceptance, which were not directly measured in this study.
Besides that, C-EQ-ALINEA could be extended in following forms:    
\begin{itemize}
    \item Enhance algorithm resilience under degraded conditions (e.g., sensor failures, communication delays, irregular congestion).
    \item Extend C-EQ-ALINEA to address multiple objectives such as emissions reduction and safety.
    \item Incorporate short-term traffic prediction and deep learning models to enable proactive congestion control and improved flow prediction.
    \item Apply on-line machine learning, for automatic adjustment of coordination gains and weight parameters using historical and real-time data.
    \item Conduct controlled field trials to validate simulations and gather real-world data on flow, safety, emissions, and user satisfaction.
\end{itemize}

%%%%%%%%%%%%%%%%%%%%%%%%%%%%%%%%%%%%%%%%%%%%%%%%%%%%%%%%%%%%%%%%%%%%%%%%%%%%%%%%
% \section{ACKNOWLEDGMENTS}

% The authors gratefully acknowledge the contribution of National Research Organisation and reviewers' comments.

%%%%%%%%%%%%%%%%%%%%%%%%%%%%%%%%%%%%%%%%%%%%%%%%%%%%%%%%%%%%%%%%%%%%%%%%%%%%%%%%

\bibliographystyle{IEEEtran}
\bibliography{references}

\end{document}